# Inverse parameter identification of strength-based cohesive zone model incorporating structural response and failure mode

Tianxiang Shi [a], Yongqiang Zhang [a,*],

[a] *College of Civil Engineering and Architecture, Zhejiang University, Hangzhou, 310058, China*

## Abstract

Composite interfaces are commonly simulated by cohesive zone model (CZM), with the key challenge being the calibration of the interfacial parameters. A recently introduced strength-based cohesive zone model (ST-CZM) exhibits parameters that significantly differ from those of traditional CZM. A new framework is presented in this paper to determine the interface parameters of ST-CZM. This approach employs the multi-island genetic algorithm (MIGA) to obtain the interface parameters aligning closely with experimental observations. The introduced framework innovatively formulates an objective function, considering both the congruence of the load-displacement curve and the alignment with the failure mode of model. A method using the interface debonding length to quantify the failure mode of the model is proposed. The inverse algorithm is then used to identify the interface parameters of both the double cantilever beam (DCB) experiment and the four-point bending test. The robustness and accuracy of the algorithm are validated through the DCB test. The findings indicate that the numerical results align closely with the experimental data, confirming that the interface parameters identified by the proposed framework can reproduce experimental results.

# 1. Introduction

At present, composite materials have emerged as pivotal constituents across diverse sectors, encompassing aerospace engineering, automotive technology, and renewable energy industry [1-3]. Their wide application and gradual substitution for traditional materials arise from the low cost, high strength, high toughness, and high specific strength. For the safety of the composite structures, it is much important to investigate their failure modes, such as delamination [4], fatigue [5], and debonding [6]. Among the various failure modes in composite structures, debonding is the most common damage mode and it is closely associated with the interface [7]. As the debonding often takes place internally, understanding the debonding mechanisms of composite structures is crucial to prevent the structural failures, and significant attention has been drawn to both experimental investigations and computational simulations concerning the interface performance. Over the past several decades, lots of cohesive zone models (CZMs) have been developed to simulate the interfacial damage. Initially proposed by Barenblatt [8] and Dugdale [9], the CZMs have been widely adopted to replace Linear Elastic Fracture Mechanics (LEFM). In the CZMs, the cohesive traction is assumed to hold the crack's upper and lower surfaces together. The propagation of the crack requires an external load to counteract the effects of this traction. This behaviour is defined by the cohesive zone (CZ) law, which describes the relationship between the interface traction and separation. Numerous CZ laws have been proposed over the years, including the bilinear [10] , exponential [11], and multilinear [12, 13] models. Recently, Nairn and Aimene [14] introduced a novel strength-based cohesive zone model (ST-CZM), offering a fresh perspective. Shi et al.[15] developed ST-CZM to the thermo-mechanical coupling. It should be noticed that the meaning in parameters of the ST-CZM is different from that in the traditional CZMs. Calibrating these new parameters in the ST-CZM requires a unique experimental method. However, the experiments are intricate and need further validation to ensure the acceptability of results.

Gong et al. [16] found that the CZ laws have a nonlinear impact on the simulation results. As such, determining the interface parameters of ST-CZM becomes an indispensable step for accurate simulations. Recent advancements have introduced inverse methods to calibrate interface parameters [12, 17, 18]. With these methods, one can derive the interface parameters inversely through comparatively straightforward experiments and simulations. The inverse methodology identifying the CZ law is operated by iteratively adjusting the interface parameters [19, 20], aiming to minimize the difference between experimental and simulated responses [21-23]. An essential step in this framework is acquiring the experimental structural responses. The most commonly measured responses are the crack mouth opening displacement (CMOD) and the applied load [19, 20, 23]. However, an objective function that solely focuses on these responses might cause simulation results to misalign with other experimental findings, such as local-level failures or general patterns of structural damage.

It is noted that many measures for local-level failure have been integrated into the identification method to enhance the reliability of the CZ laws. For instance, the digital image correlation (DIC) has been utilized to capture data on crack tip opening displacement [24, 25]. Additionally, the inverse algorithms have been formulated considering both global and local responses. Xu et al. [26] considered the combined effects of DIC measurement data and load-displacement curves. The embedded fiber Bragg grating (FBG) sensors are utilized to record distributed strain along the delamination extension path, introducing an new objective function for inverse parameterization [27]. In the works of considering multiple responses, the cohesive zone laws are identified through minimizing such multi-objective problems. However, such methods demand advanced experimental measurement techniques, and there is no certainty that the CZ laws, ascertained through inverse identification, will precisely emulate the failure mode of experiment.

Even minor variations in parameters can induce substantial differences in the delamination predictions during mixed-mode fracture, resulting in entirely different failure mechanisms [28]. It has been proved by in-situ experiment [29] that the mixed-mode condition has a significant impact on the damage and failure process of the interface. Wan et al. [30] effectively illustrated a correlation between the debonding length and the structural response of interface through both the 3D Discrete Element Method (DEM) and the experiment. Fu et al. [31] proved that the length and position of interface debonding influence the resultant failure modes in the experiment. Based on previous work, this study quantifies the failure mode by assessing the interface debonding length and integrates it into the objective function. It is a new perspective for inverse identification.

There are various choices for parameter optimization methods. Jensen et al. [13] introduced a method for n-segmented multilinear cohesive laws, employing a gradient-based optimization algorithm for their inverse identification. Cao et al. [32] presented an inverse method based on the Euler-Bernoulli beam theory to identify the cohesive zone laws under mode I, simultaneously confirming the method's robustness.

On the other hand, the traditional genetic algorithms (GA) is employed for inverse analysis to determine the CZ laws [33], comparing the numerical results with the experiment data on mixed-mode delamination of carbon epoxy materials.

It is a novel approach using GA to optimize interface parameters in inverse problems. However, many previous methods rely on the traditional GA, which can easily become trapped in local optima, particularly when solving highly nonlinear problems. As an improvement over traditional methods, the Migration Genetic Algorithm (MIGA) effectively avoids local optima issues. The superiority of the MIGA algorithm has been acknowledged in several studies. For instance, MIGA was utilized to optimize the recycling process parameters of carbon fibre [34], while others employed MIGA for structural design optimization [35, 36].

Utilizing the MIGA algorithm, this paper presents a novel framework for the inverse identification of crucial parameters in ST-CZM. We introduce a pioneering method that estimates structural failure modes based on debonding length while considering the similarity of load-displacement curves, which is achieved by a nonlinear objective function. Additionally, a comprehensive numerical framework is proposed, facilitating parameter inversion for a wide range of interface models.

## 2. The strength theory based cohesive zone model (ST-CZM)

Recently, a new cohesive zone model based on strength theory is proposed by Nairn and Aimene [14], which redefines the traction-separation law as a strength model that closely align with the continuum damage mechanics. Fig.1 depicts a sawtooth cohesive law, which can be expressed as

$$S_i(\omega_i) = \sigma_{i,c} \frac{\Delta u_{i,f} - \omega_i}{\Delta u_{i,f} - \Delta u_{i,c}}, i = n, t \tag{1}$$

Herein, the subscripts n and t denote the normal and tangential directions, respectively. The term $\sigma_{i,c}$ represents the peak cohesive traction, while $k_{i,0}$ stands for the initial elastic stiffness. Furthermore, $\Delta u_{i,c}$ and $\Delta u_{i,f}$ correspond to the displacement gaps related to damage initiation and vanishment of traction, respectively. In the elastic phase ($\Delta u_i \leq \Delta u_{i,c}$), the damage state variable $\omega_i$ corresponds to $\Delta u_{i,c}$ in the ST-CZM. However, once the damage is evaluated, $\omega_i$ changes with the maximal displacement gap ($|\Delta u_{i,\max}|$).

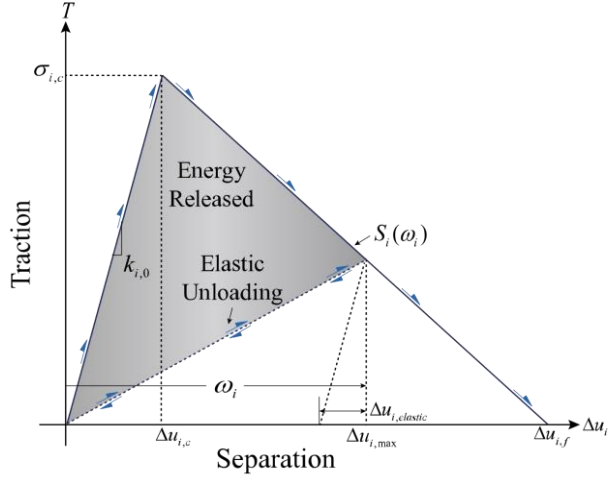

Fig.1 A bilinear cohesive traction-separation curve

Utilizing Eq.(1), the tractions and the associated strain energy can be expressed as

$$T_i = \frac{S_i(\omega_i)}{\omega_i}\Delta u_i \qquad (2)$$

$$U(\Delta u_i, \omega) = \frac{1}{2}\frac{S_i(\omega_i)}{\omega_i}\Delta u_i^2, \ U(\Delta \bar{\mathbf{u}}, \omega) = U_n + U_t \qquad (3)$$

where $\Delta\bar{\mathbf{u}} = \begin{bmatrix} \Delta u_n & \Delta u_t \end{bmatrix}$ is the displacement gap vector.

The damage scalars $D_n$ and $D_t$ are

$$D_n = 1 - \frac{S_n(\omega_n)}{k_{n,0}\omega_n}, \ D_t = 1 - \frac{S_t(\omega_t)}{k_{t,0}\omega_t} \qquad (4)$$

Using a second-rank damage tensor $\mathbf{D} = \mathrm{diag}(D_n, D_t)$, Eqs.(2) and (3) can be reformulated as

$$\mathbf{T} = \mathbf{k}_0(\mathbf{I}-\mathbf{D})\Delta\bar{\mathbf{u}} \qquad (5)$$

$$U(\Delta\bar{\mathbf{u}}, \mathrm{D}) = \frac{1}{2}\mathbf{k}_0(\mathrm{I}-\mathrm{D})\Delta\bar{\mathbf{u}}\cdot\Delta\bar{\mathbf{u}} \qquad (6)$$

where $\mathbf{k}_0$ is the initial stiffness.

For mixed-mode failure scenarios, use is made of an elliptical failure surface to integrate both the normal and tangential tractions, as described by

$$\left(\frac{\langle T_n \rangle}{S_n(\omega_n)}\right)^2 + \left(\frac{T_t}{S_t(\omega_t)}\right)^2 = 1 \qquad (7)$$

Here, the McAuley bracket $\langle x \rangle$, denoted as $\max(0, x)$, is used to eliminate the compressive traction. The aggregate traction reaches the failure surface when $\|\Delta\mathbf{u}\| \to \omega_d$. The combined damage state variable $\omega_d$ is written as

$$\omega_d = \left( \frac{\sin^2 \gamma}{\omega_n^2} + \frac{\cos^2 \gamma}{\omega_t^2} \right)^{-\frac{1}{2}} \tag{8}$$

where $\gamma$ is defined as the mode-mixty ( i.e. $\tan \gamma = \Delta u_n / \Delta u_t$ ).

Eq.(7) suggests that as the damage state variables $\omega_n$ and $\omega_t$ evolve, the failure surface contracts. To monitor the failure surface, Nairn and Aimene [14] introduced the trial traction. Through identifying the state of the trial traction, one can update the failure surface by

$$\left( \frac{T_n (\Delta u_n + d \Delta u_n)}{S_n (\omega_n + d \omega_n)} \right)^2 + \left( \frac{T_t (\Delta u_t + d \Delta u_t)}{S_t (\omega_t + d \omega_t)} \right)^2 = \left( \frac{\Delta u_n + d \Delta u_n}{\omega_n + d \omega_n} \right)^2 + \left( \frac{\Delta u_t + d \Delta u_t}{\omega_t + d \omega_t} \right)^2 = 1 \tag{9}$$

Expanding Eq.(9) with a Taylor series and retaining only the first-order terms yields the equation for the evolution of the damage state variables, expressed as

$$\frac{\Delta u_n^2}{\omega_n^3} d\omega_n + \frac{\Delta u_t^2}{\omega_t^3} d\omega_t = \frac{\Delta u_n}{\omega_n^2} d \Delta u_n + \frac{\Delta u_t}{\omega_t^2} d \Delta u_t \tag{10}$$

Assuming that $D_n = D_t = D$ the damage update can be expressed as

$$dD = \frac{\dfrac{\Delta u_n}{\omega_n^2} d \Delta u_n + \dfrac{\Delta u_t}{\omega_t^2} d \Delta u_t}{\dfrac{1}{R_n(\omega_n)} \dfrac{\Delta u_n^2}{\omega_n^3} + \dfrac{1}{R_t(\omega_t)} \dfrac{\Delta u_t^2}{\omega_t^3}} \tag{11}$$

with

$$R_i(\omega_i) = \frac{dD}{d\omega_i} = \frac{\varphi_i(\omega_i)}{k_{i,0} \omega_i^2} \tag{12a}$$

$$\varphi_i(\omega_i) = S_i(\omega_i) - \omega_i \frac{d S_i(\omega_i)}{d \omega_i} \tag{12b}$$

Through Eqs.(11)-(12), the damage in both normal and tangential directions is initialized and reaches unity at the same time. Additionally, these equations enable us to determine the damage evolution and the tangential stiffness matrix of the element. We can calibrate the key parameters for ST-CZM through the derivation: the peak traction $\sigma_{i,c}$, the damage initiation displacement $\Delta u_{i,f}$, and the failure displacement $\Delta u_{i,c}$.

Contrary to the traction-separation law, the ST-CZM introduces a redefined cohesive zone model based on the strength concept. Up to now, the prior research is lack of guidance on the selection of its specific parameters. Although experimental calibration is the most common approach, it demands extensive testing and can be prohibitively expensive. This paper presents a new framework for identifying the parameters of the ST-CZM to solve this problem.

## 3. Methodology

### 3.1 Proposed optimization algorithm

The GA originates from the interdisciplinary fields of biology and artificial intelligence and emerges as an efficient global optimization technique. It has been extensively used in structural design processes because it is not bound by the continuity of design variables and does not depend on gradient information [37-39].

However, the traditional GA often gets trapped in local optima, which limits their ability to effectively solve complex nonlinear optimization problems. To enhance the performance of GAs, various types of new algorithms have been developed [40, 41]. The Multi- Island Genetic Algorithm (MIGA) is introduced in this context. The primary distinction between MIGA and traditional GAs is that the MIGA segments the population into multiple islands, enabling chromosomes to migrate between these islands (as shown in Fig.2). The migration feature ensures that the population of each island continuously receives new genetic information, resulting in enhanced search performance by the algorithm. The MIGA can escape local optima and attain global nondominated solutions, improving the reliability of the optimal solution [36, 42].

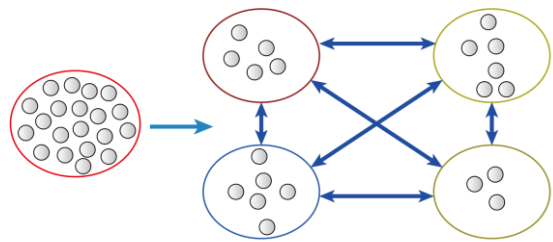

Fig.2 debonding growth and dissipated energy evolution during monotonic opening

The procedure of the MIGA is summarized as follows:

Step 1): Initialize the global variables (i.e. initial guess for interfacial parameters and hyper-parameters for MIGA), then divide the population into multiple islands and distribute the global variables to each island;

Step 2): Each island operates in parallel, undergoing traditional genetic processes such as selection,

recombination, mutation, and migration;

Step 3): Receive and update the optimal solution for each island until the optimization objective is met;

Until now, the majority of inversion methods only focus on the load-displacement curves of simulation results[17, 22, 28]. In the complex modelling, however, these methods are always unilateral. For example, when not constrained by stringent boundary conditions, the framework may result in significantly different failure modes for specimens, even if the identification process sometimes yields accurate load-displacement curves. Thus, it is necessary to have a more accurate estimation of the interface parameters to solve this problem, which is difficult in the laboratory and engineering practice.

This paper introduces a novel framework that employs MIGA for parameter optimization in the cohesive zone model based on strength theory. The framework facilitates the optimal alignment of numerical results with experimental results by determining the appropriate parameters. As the interface debonding is often regarded as a representation of the specimen's failure mode in many experiments [30, 31, 43], the proposed framework employs the debonding length of interface to characterize the failure mode of specimen. This method involves extracting the total length of the damaged element from UEL, which is a technique that has not been proposed in inverse methods before. This innovative framework not only aligns with the load-displacement curves, but also predicts the failure mode of computational structures, overcoming a shortcoming inherent of traditional frameworks.

The workflow of the proposed inversion framework is illustrated in Fig.3, with the main steps detailed as follows:

Step 1): Provide an initial guess of the interface parameters;

Step 2): Optimize the parameters using the MIGA;

Step 3): Employ the UEL to compute the strength model with the given input parameters;

Step 4): Extract load-displacement curves and debonding length-displacement curves from numerical results using Python scripts;

Step 5): Calculate the objective function (further discussed below);

Step 6): Evaluate whether the predicted results satisfy the convergence criteria. If not, return to Step 2;

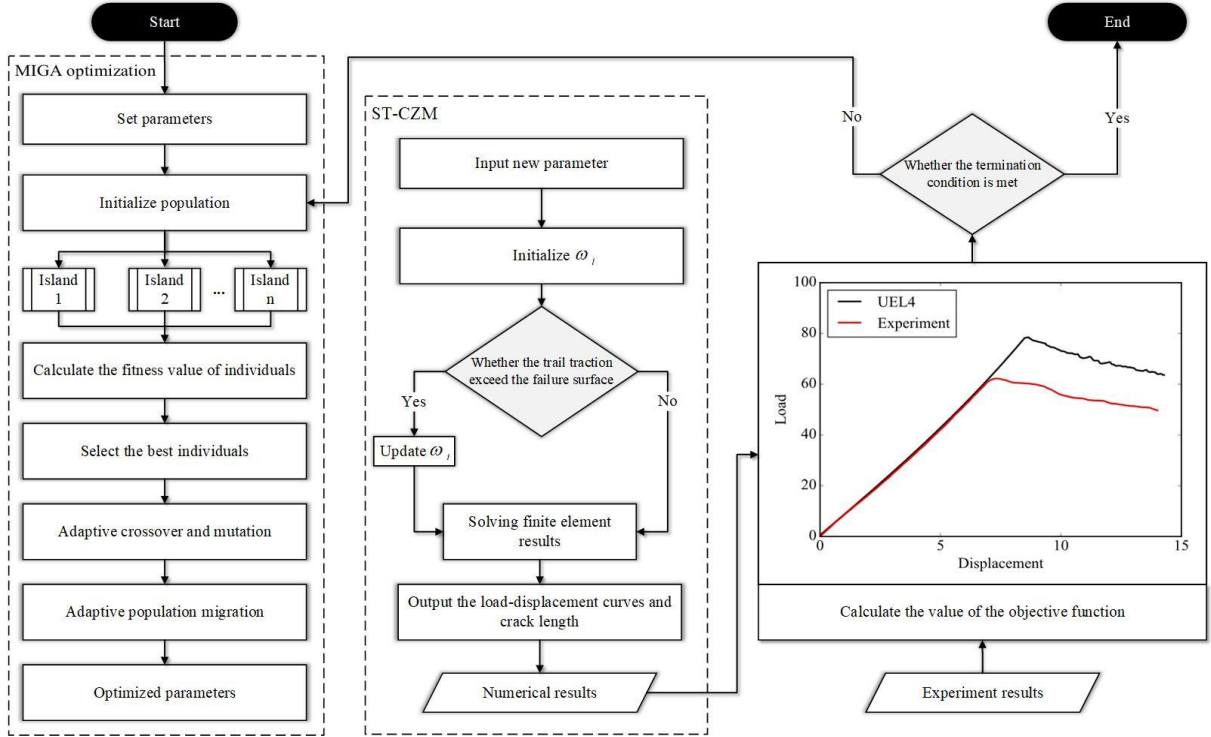

Fig.3 The workflow of the proposed framework

Although we utilize the proposed framework to identify the parameters of ST-CZM, it can also be used to identify parameters for any other constitutive model implemented in UEL, due to an application program interface (API) having been developed.

## 3.2 Objective function

To start the identification framework, an initial estimate of the interface parameters is required (as shown in Fig.3). The key parameters of the ST-CZM are introduced as $\sigma_{i,c}$, $\Delta u_{i,f}$ and $\Delta u_{i,c}$. In order to construct the objective function $f(x)$ which is discussed blow, the design variables are parametrized as

$$\mathbf{x} = \begin{bmatrix} \sigma_{n,c} & \Delta u_{n,f} & \Delta u_{n,c} & \sigma_{t,c} & \Delta u_{t,f} & \Delta u_{t,c} \end{bmatrix}^T \quad (13)$$

The proposed framework will yield the optimal parameter that matches the experimental results. Once these parameters are determined, the associated fracture energy and initial stiffness can be found out (using Eqs.(3) and (6)). Meanwhile, the ST-CZM also needs to update the damage state variables based on these parameters.

The least-squares norm is used to formulate the objective function, quantifying the difference between the experimental results and the corresponding simulation outcomes. In the current framework, the objective function is nonlinear, and is defined as

$$f(x) = \alpha_L \|\varphi_L(x)\|^2 + \alpha_F \|\varphi_F(x)\|^2 \tag{14}$$

Here, the values of scalars $\alpha_L$ and $\alpha_F$ vary between 0 and 1, and they serve to weight the two addends, ensuring their comparability at the beginning of the inverse parameter identification.

The objective function considering the debonding length is written as:

$$\varphi_L(x) = \frac{\mathbf{L}^{\exp} - \mathbf{L}^{nu}(x)}{\|\mathbf{L}^{\exp}\|} = \frac{\mathbf{r}_L(x)}{\|\mathbf{L}^{\exp}\|} \tag{15}$$

where

$$\mathbf{r}_L(x) = \begin{Bmatrix} L_1^{\exp} - L_1^{nu}(x) \\ L_2^{\exp} - L_2^{nu}(x) \\ \vdots \\ L_n^{\exp} - L_n^{nu}(x) \end{Bmatrix} \tag{16}$$

Eq.(15) illustrates the normalization of the discrepancy in interfacial debonding length between experimental and simulated results of a specific measuring point. $L_n^{\exp}$ denotes the debonding length obtained from experiment, and $L_n^{nu}(x)$ is determined by calculating the total length of the cohesive elements for which the damage scalar $D$ is 1. It should be noted that the subscript $n$ represents the computational step associated with a specific displacement value.

The objective function, accounting for variations in the load-displacement curves, is expressed as

$$\varphi_F(x) = \frac{\mathbf{F}^{\exp} - \mathbf{F}^{nu}(x)}{\|\mathbf{F}^{\exp}\|} = \frac{\mathbf{r}_F(x)}{\|\mathbf{F}^{\exp}\|} \tag{17}$$

where the residual vector $\mathbf{r}_F(x)$ is given by

$$\mathbf{r}_F(x) = \begin{Bmatrix} F_1^{\exp} - F_1^{nu}(x) \\ F_2^{\exp} - F_2^{nu}(x) \\ \vdots \\ F_n^{\exp} - F_n^{nu}(x) \end{Bmatrix} \tag{18}$$

Eq.(17) denotes the normalization of the load difference corresponding to a specific measuring point, where $F_n^{\exp}$ is the load obtained from experiment, and $F_n^{nu}(x)$ is the load obtained from numerical results which is output from Abaqus. It's worth noting that, owing to the characteristic of the finite element method, convergence step sizes are always changed. To streamline the calculation of objective function, both simulation and test curve data are processed using interpolation techniques, ensuring a one-to-one correspondence in the results.

## 4. Results and analysis

In this section, we establish a finite element model that mirrors the experimental setup within the proposed framework. Initially, the double cantilever beam (DCB) test is employed to validate the efficacy, precision, and robustness of our proposed inversion method. Subsequently, the framework is extended to more complex models (a full-scale concrete beam reinforced with fibre reinforced polymer (FRP)).

### 4.1 Case study: DCB test

The material properties of the DCB model are detailed in Table 1. The geometric characteristic and boundary conditions of the DCB model are [30]: the length $l = 180\text{mm}$, the width $w = 20\text{mm}$, the initial length of the crack $a_0 = 50\text{mm}$, and the thickness $2h = 3.24\text{mm}$ (as shown in the Fig.4). The cohesive elements are implemented by UEL in Abaqus, with the specimen modelled by solid elements. The interface parameters are identified by the proposed framework. To ensure the accuracy in determining the bonding length of interface, a dense mesh (0.1mm per element) is employed in the numerical model.

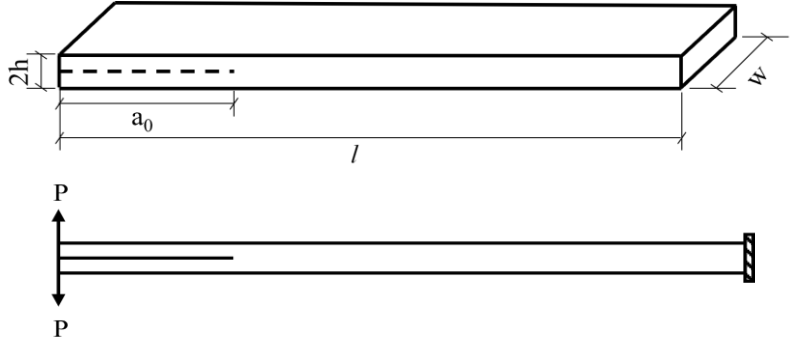

Fig.4 Geometry properties and boundary conditions of double cantilever beam specimen

In the DCB model, the crack propagation is linear. Python scripts output the number and position of the damage interface elements, which are used to calculate the crack length.

Table 1 Material properties for the DCB model

| | $E_{11}$[GPa] | $E_{22} = E_{33}$[GPa] | $G_{12} = G_{23} = G_{33}$[MPa] | $v_{12} = v_{13}$ | $v_{23}$ |
|---|---|---|---|---|---|
| Solid Elements | 130 | 6.5 | 2.7 | 0.26 | 0.50 |

Initially, a set of interface parameters (detailed in Table 2) is selected at random. The numerical outcomes based on these parameters are illustrated in Fig.5.

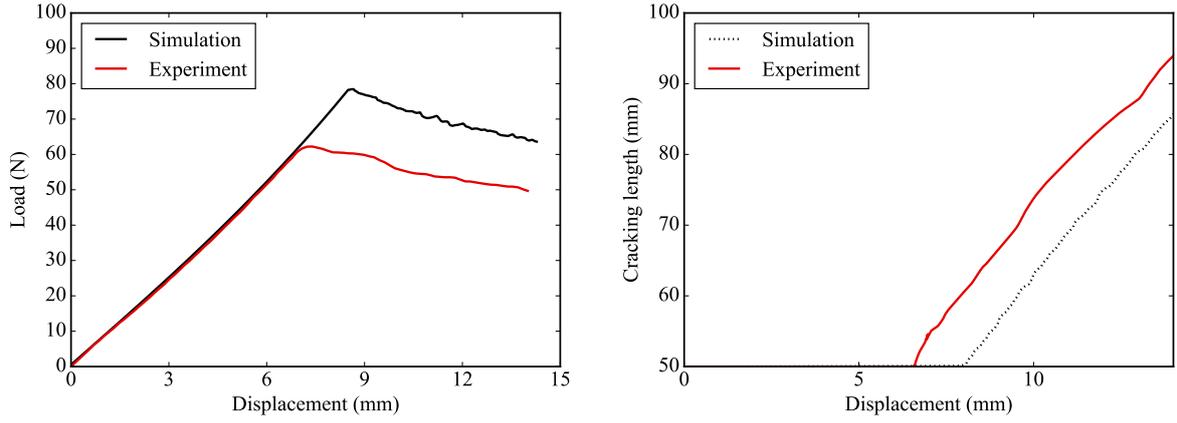

(a) Load-displacement curves  (b) Debonding length-displacement curves
Fig.5 Comparation between simulation and experiment results (before MIGA optimization)

Table 2 Initial guess of interface parameters

| | $\sigma_{n,c}$ [MPa] | $\sigma_{t,c}$ [MPa] | $\Delta u_{n,f}$ [mm] | $\Delta u_{t,f}$ [mm] | $\Delta u_{n,c}$ [mm] |
|---|---|---|---|---|---|
| Cohesive Element | 40 | 60 | 0.0003 | 0.000722 | 0.003 |
| | $\Delta u_{t,c}$ [mm] | $k_{i,0}$ [MPa/mm] | | | |
| | 0.003 | 133333.333 | | | |

The numerical results are completely different from the experimental findings. Notably, the peak traction from the FEM exceeds the experimental value, and the simulated debonding length significantly larger than experimental observations. This discrepancy emphasizes the significant influence of interface parameters on simulation outcomes, highlighting the importance for appropriate parameter selection.

Using the proposed inversion framework, the interface parameters are re-evaluated. Upon inputting the initial interfacial parameters and numerical results, the inverse identification method is initiated. After 200 iterations, a convergence is observed. Fig.6 depicts the iterative process of the inverse algorithm, with the objective function stabilizing near zero, and the result can be considered as the optimal solution.

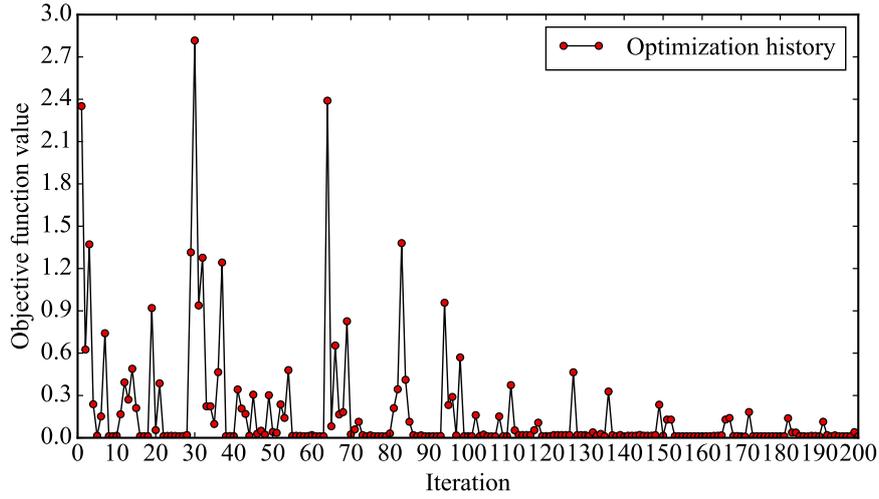

Fig.6 The iteration process of MIGA

It can be observed that in the early stage of iteration, there is notable fluctuation in the objective function, with many results not meeting the expectations. After 100 iterations, the numerical outcomes, derived from inversion parameters, start to stabilize within a permissible range, bringing the objective function value near zero. The optimal interface parameter values obtained by the proposed framework are presented in Table 3.

Table 3 Results of inverse identification for interface parameters

|  | $\sigma_{n,c}$ [MPa] | $\sigma_{t,c}$ [MPa] | $\Delta u_{n,f}$ [mm] | $\Delta u_{t,f}$ [mm] | $\Delta u_{n,c}$ [mm] |
|---|---|---|---|---|---|
| Cohesive Element | 27 | 40 | 0.00027 | 0.00025 | 0.003273 |
|  | $\Delta u_{t,c}$ [mm] | $k_{n,0}$ [MPa/mm] | $k_{t,0}$ [MPa/mm] |  |  |
|  | 0.003273 | 100000 | 160000 |  |  |

Numerical results based on the optimized parameters are depicted in Fig.7. The load-displacement curves from both the experiment and simulation align closely, and the variation trend in simulated bonding length mirrors the experimental observations. The findings demonstrate that the inverse identification method accurately predicts both the load-displacement outcomes and the failure mode.

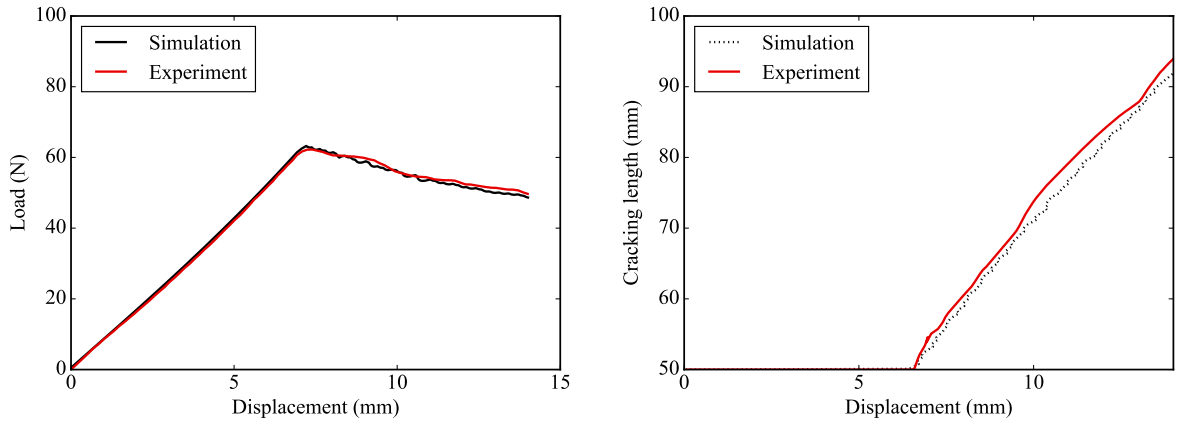

(a) Load-displacement curves　　　　　　(b) Debonding length-displacement curves

Fig.7 Comparation between simulation and experiment results (after MIGA optimization)

## 4.2 Comparation with NSGA-II

To validate the superior performance of MIGA chosen for this study, we replace MIGA with the Non-dominated Sorting Genetic Algorithm-II (NSGA-II)[44] and discuss the difference in this section.

NSGA-II is a non-dominant, multi-objective optimization genetic algorithm rooted in Pareto optimal solutions and incorporates an elite retention strategy [32]. It primarily consists of the following three computation steps:

Step 1): Conduct rapid non-dominated sorting of the Pareto solution set;

Step 2): Compute crowding distances value and implement crowded-comparison operators;

Step 3): Execute crossover and mutation operations, ensuring the survival of dominant solutions by employing the non-dominated sorting mechanism;

Through these, the best solutions are retained and the optimal parameter points are identified. Repeat the above steps until the target number of iterations is met or convergence conditions are achieved.

It can be observed in Fig.8 that while NSGA-II converges more rapidly within the first 200 generations, it converges to a local optimum with the objective function around the value of 0.1.

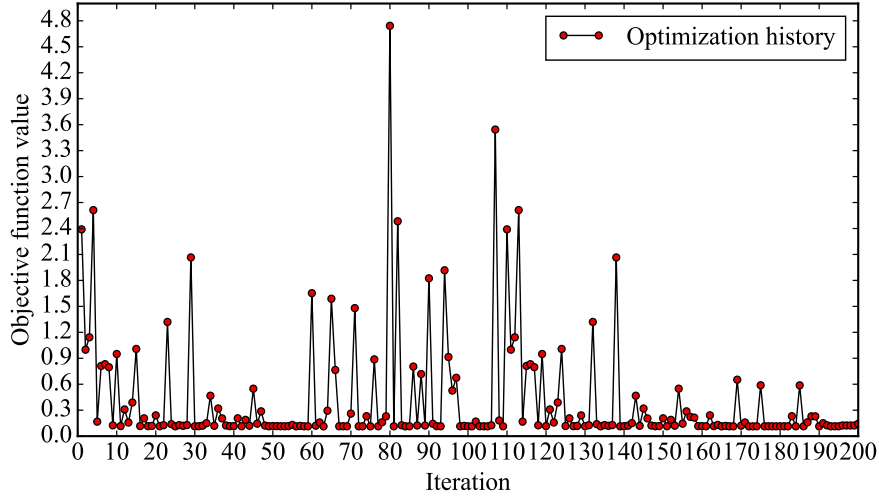

Fig.8 The iteration process of NSGA-II

Table 4 displays the interface parameters derived from the NSGA-II optimization. Notably, the tangential peak traction acquired by NSGA-II is smaller than that from MIGA, and the predicted initial stiffness is also lower. Although the load-displacement results (illustrated in Fig.9(a)) are in an acceptable error margin, the debonding length-displacement curves show discrepancies (refer to Fig.9(b)).

Table 4 Results of inverse identification for interface parameters

| Cohesive Element | $\sigma_{n,c}$ [MPa] | $\sigma_{t,c}$ [MPa] | $\Delta u_{n,f}$ [mm] | $\Delta u_{t,f}$ [mm] | $\Delta u_{n,c}$ [mm] |
|---|---|---|---|---|---|
| | 24 | 28 | 0.00029 | 0.00025 | 0.00375 |
| | $\Delta u_{t,c}$ [mm] | $k_{n,0}$ [MPa/mm] | $k_{t,0}$ [MPa/mm] | | |
| | 0.00350 | 82758.62 | 112000 | | |

The results indicate that when implementing the inversion framework from this study using NSGA-II, the derived solutions tend to the local optima. When applying this method to complex models, such mismatches can result in significantly different failure mode from experiment. The comparison demonstrates that the MIGA is more suitable to present the inversion framework, holding promising potential for interface parameter inversion in intricate models.

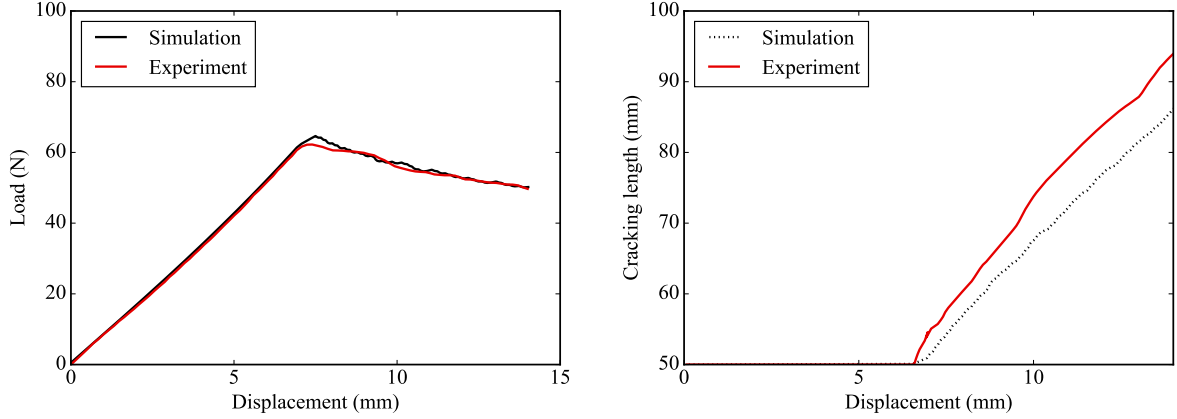

(a) Load-displacement curves        (b) Debonding length-displacement curves

Fig.9 Comparation between simulation and experiment results (after NSGA-II optimization)

## 4.3 Robustness and accuracy

Existing literature consistently acknowledges that the inverse problems are ill-posed, which means that the solutions are not unique [26]. This suggests that there are multiple local minima in the feasible domain. Since the solution is not unique in inverse problems, the validation of the robustness and accuracy of the proposed framework is necessary. To discuss the potential local minima that may arise in optimization, this section investigates the results based on different initial parameter guesses. Table 5 lists four varied sets of initial parameters.

Table 5 The initial guess values for ST-CZM1 to ST-CZM4

|  | $\sigma_{n,c}$ [MPa] | $\sigma_{t,c}$ [MPa] | $\Delta u_{n,f}$ [mm] | $\Delta u_{t,f}$ [mm] | $\Delta u_{n,c}$ [mm] | $\Delta u_{t,c}$ [mm] | $k_{n,0}$ [MPa/mm] | $k_{t,0}$ [MPa/mm] |
|---|---|---|---|---|---|---|---|---|
| ST-CZM1 | 20 | 26 | 0.0003 | 0.0003 | 0.006273 | 0.006273 | 66666.67 | 86666.67 |
| ST-CZM2 | 20 | 26 | 0.0003 | 0.0003 | 0.004273 | 0.004273 | 66666.67 | 86666.67 |
| ST-CZM3 | 20 | 40 | 0.00025 | 0.00025 | 0.004273 | 0.004273 | 80000.00 | 160000.00 |
| ST-CZM4 | 40 | 50 | 0.0003 | 0.0003 | 0.002573 | 0.002573 | 133333.33 | 166666.67 |

The first set features the largest failure displacement and fracture energy. The second set has a reduced fracture energy compared with the first set while it exhibits the simulation results which is close to experimental findings. Compared with the second set, the third set increases the peak traction and decreases the damage initiation displacement. Based on the third set, the fourth set increases the peak traction and damage initial displacement in both normal and tangential directions, and decreases the failure displacement.

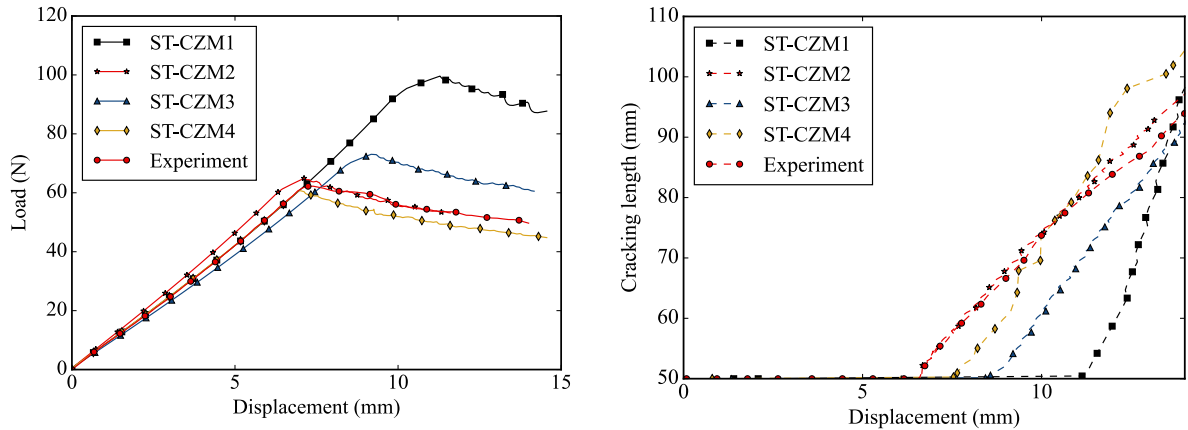

(a) Load-displacement curves  (b) Debonding length-displacement curves
Fig.10 Comparation between simulation and experiment results (before MIGA optimization)

The load-displacement curves corresponding to the four sets of initial guesses are depicted in Fig.10(a), and the debonding length-displacement curves are shown in Fig.10(b).

The results obtained by the first set display the highest peak load and initial failure displacement, significantly different from the experimental findings. In contrast, the results simulated by the second set (including both load-displacement and debonding length-displacement curves) align more closely with experiments. However, it is not the optimal result. The outcomes from the third set slightly exceed the experimental findings in the peak traction and the initial crack growth displacement. The fourth set predicts lowest peak load and initial debonding length. The comparison of the simulation results indicates that the fracture energy (using Eq.(3)), has a significant influence on the simulation.

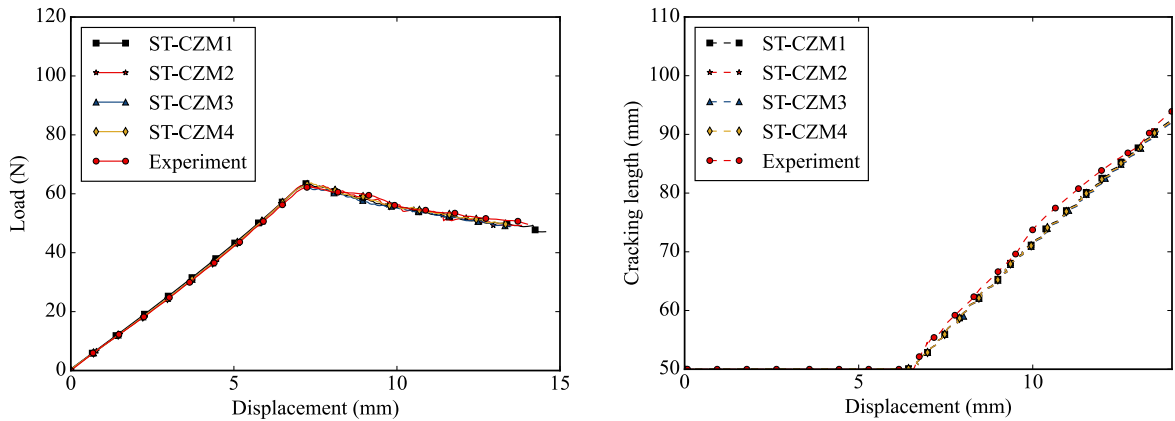

(a) Load-displacement curves  (b) Debonding length-displacement curves
Fig.11 Comparation between simulation and experiment results (after MIGA optimization)

To demonstrate the robustness of the inverse framework, numerous decimal places are added to each parameter during the selection of initial parameters, ensuring that each group has distinct characteristic. The results optimized by the proposed framework are presented in Table 6. It can be observed that the optimization results are slightly different from one another, stemming from the fact that the optimal solution for a nonlinear problem is not unique. Considering that the minor differences beyond a few decimal places will not significantly impact the simulation, the results are deemed acceptable. The simulated results are shown in Fig.11. Consequently, the optimization framework has demonstrated robustness, and it is insensitivity to the selection of initial guesses.

Table 6 Results from inverse parameter identification for ST-CZM1 to ST-CZM4

|  | $\sigma_{n,c}$ [MPa] | $\sigma_{t,c}$ [MPa] | $\Delta u_{n,f}$ [mm] | $\Delta u_{t,f}$ [mm] | $\Delta u_{n,c}$ [mm] | $\Delta u_{t,c}$ [mm] | $k_{n,0}$ [MPa/mm] | $k_{t,0}$ [MPa/mm] |
|---|---|---|---|---|---|---|---|---|
| ST-CZM1 | 27 | 40 | 0.0002833 | 0.0002739 | 0.003299 | 0.003295 | 95305.33 | 146038.70 |
| ST-CZM2 | 27 | 39 | 0.0002638 | 0.0002785 | 0.003174 | 0.003229 | 102350.27 | 132854.58 |
| ST-CZM3 | 27 | 40 | 0.0002833 | 0.0002731 | 0.003225 | 0.003206 | 95305.33 | 146466.50 |
| ST-CZM4 | 27 | 38 | 0.0002800 | 0.0002980 | 0.003232 | 0.003206 | 96428.57 | 127516.78 |

We have chosen the optimization history of the first group to illustrate the optimization process (as shown in Fig.12). It can be observed that around the 120th iterative step, the objective function converges to zero. The stress contour plots for 45th iteration and 200th iterations are also shown in Fig.12.

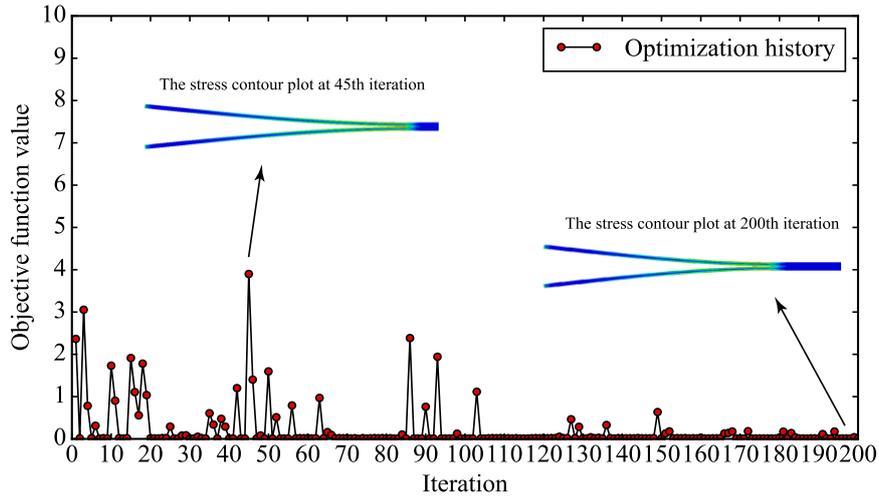

Fig.12 The iteration process and stress contour plot at 45th and 200th iteration

Fig.13 depicts both the load-displacement curve and the debonding length-displacement curve from the 45th step of the ST-CZM1 optimization process. The interface parameters corresponding to this step are presented in Table 7.

Table 7 Results from inverse parameter identification

| | $\sigma_{n,c}$ [MPa] | $\sigma_{t,c}$ [MPa] | $\Delta u_{n,f}$ [mm] | $\Delta u_{t,f}$ [mm] | $\Delta u_{n,c}$ [mm] |
|---|---|---|---|---|---|
| Cohesive Element | 21 | 36 | 0.00028 | 0.00029 | 0.00375 |
| | $\Delta u_{t,c}$ [mm] | $k_{n,0}$ [MPa/mm] | $k_{t,0}$ [MPa/mm] | | |
| | 0.00350 | 75000 | 124137 | | |

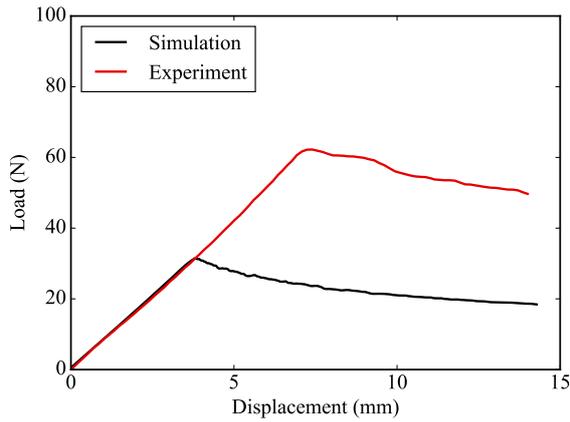 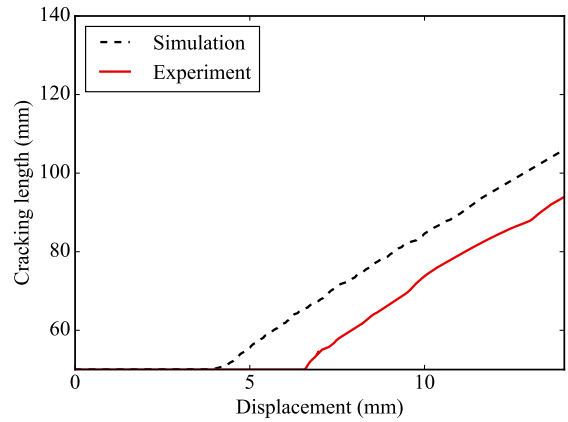

(a) Load-displacement curves　　　　　(b) Debonding length-displacement curves
Fig.13 Comparation between simulation and experiment results (at 45th iteration)

It can be observed in Fig.13 that when simulating with the intermediate parameters during the optimization process, both the peak traction and the initial debonding length exceed the experimental results. The simulation results obtained by optimal parameters are shown in Fig.14. By comparing two numerical results, it can be seen that the accuracy of intermediate results cannot be guaranteed. Hence, employing the proposed framework requires sufficient iterations to identify optimal results that satisfy the objective function. Once the stable convergence of the objective function is obtained, the results are credible.

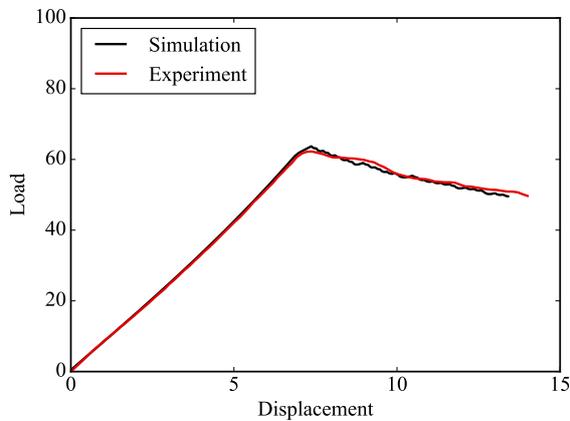 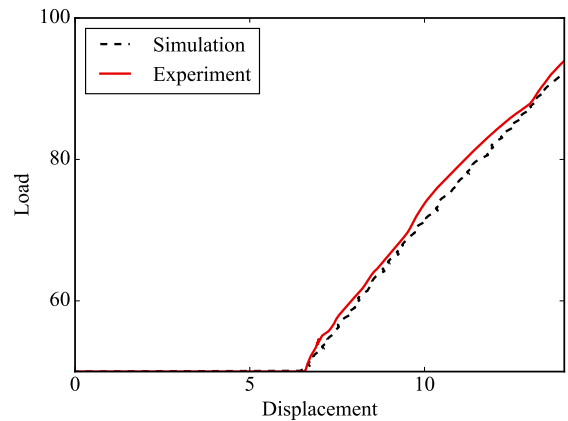

(a) Load-displacement curves　　　　　(b) Cracking length-displacement curves
Fig.14 Comparation between simulation and experiment results (at 200th iteration)

### 4.4 Case study: the four-point bending test

The failure mode of the DCB test is relatively straightforward. To validate that the parameters derived from the proposed inverse framework can faithfully reproduce the failure mode of experiment, this section provides another verification using a comprehensive experimental model [31] (i.e. the four-point bending test).

The geometric properties and boundary conditions of the test are shown in Fig.15. The steel bars are embedded in the concrete. The compressive strength of the concrete is 47 MPa, and the large-span concrete beam measures a length of $L_{beam}=4000\text{mm}$, a width of $w_{beam}=200\text{mm}$, and a height of $H=450\text{mm}$. The reinforced bars are detailed in Fig.15. The FRP plate, adhered to the bottom, has a length of $L_{FRP}=3800\text{mm}$, a width of $w_{FRP}=100\text{mm}$, and a thickness of $t_{FRP}=0.999\text{mm}$. $a$ is defined as the distance between the two-loading point.

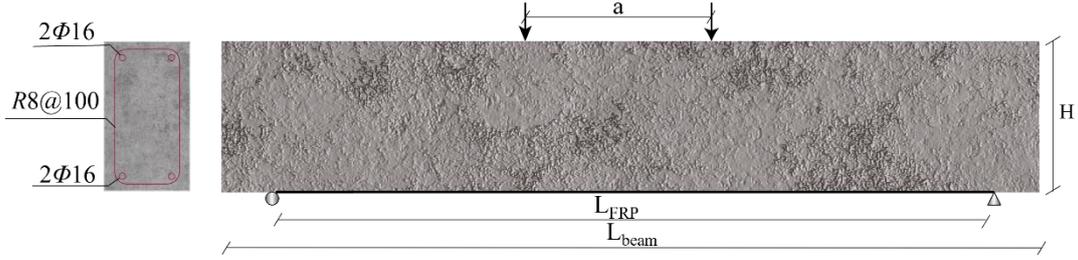

Fig.15 Geometry of the test specimen

In the experiment [31], the value of the debonding length across complete process is not provided. However, the debonding length corresponding to both initial and final debonding are given, which can also form the objective function. For this experiment, the objective function is simplified to

$$\mathbf{r}_L(\mathbf{x}) = \begin{pmatrix} L_1^{\exp} - L_1^{nu}(\mathbf{x}) \\ L_{end}^{\exp} - L_{end}^{nu}(\mathbf{x}) \end{pmatrix}$$ without intermediate variable.

The debonding length is estimated to be 436mm at the onset of local debonding. At the end of the experiment, the strength contribution from the FRP plate has nearly vanished, with the final debonding length approximated at 1048mm[31]. To compare the simulation results with the experimental data, the load is converted to the mid-span moment. This adjustment does not affect the objective function.

Fig.16 depicts the iteration process during the optimization, the objective function eventually converging around zero. Table 8 details the interface parameters derived from the proposed framework.

Table 8 Results from inverse parameter identification

| Cohesive Element | $\sigma_{n,c}$ [MPa] | $\sigma_{t,c}$ [MPa] | $\Delta u_{n,f}$ [mm] | $\Delta u_{t,f}$ [mm] | $\Delta u_{n,c}$ [mm] | $\Delta u_{t,c}$ [mm] |
|---|---|---|---|---|---|---|
| | 12 | 14 | 0.02513 | 0.02897 | 0.1457 | 0.1504 |

Fig.17 illustrates the load-displacement curve for this simulation. It can be found that even though the objective function is simplified, the results are in the acceptable range.

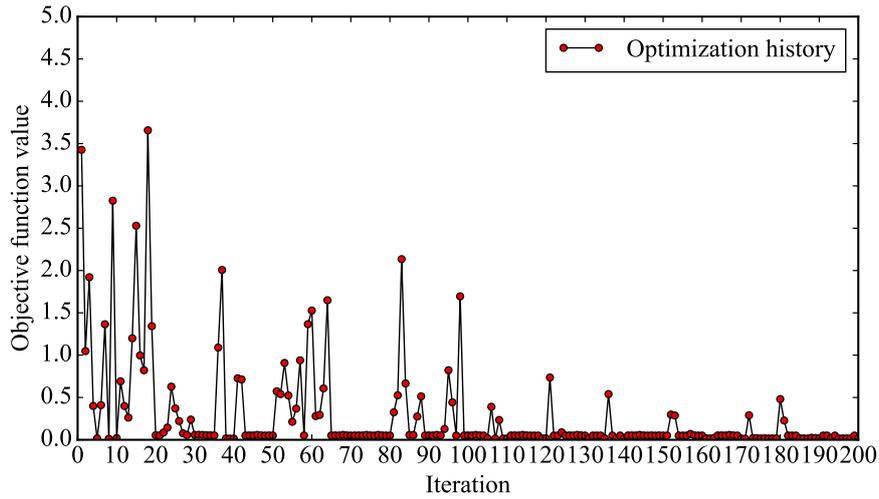
Fig.16 The iteration process

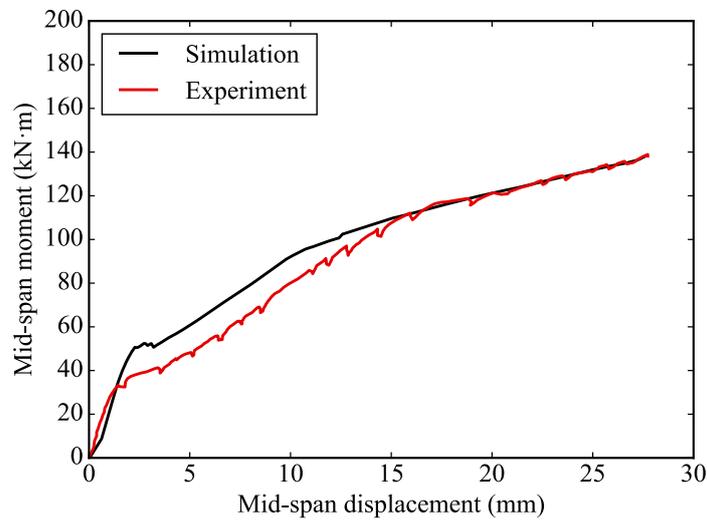
Fig.17 Moment-deflection curves

Fig.18 illustrates the failure mode of experiment, while Fig.19 presents the failure mode obtained by simulation. It can be observed that the numerical results, derived from optimal parameters, mirror the failure mode in the experiment. This result signifies a notable advancement from previous inversion methods which lacked such precision in replicating experimental failure modes.

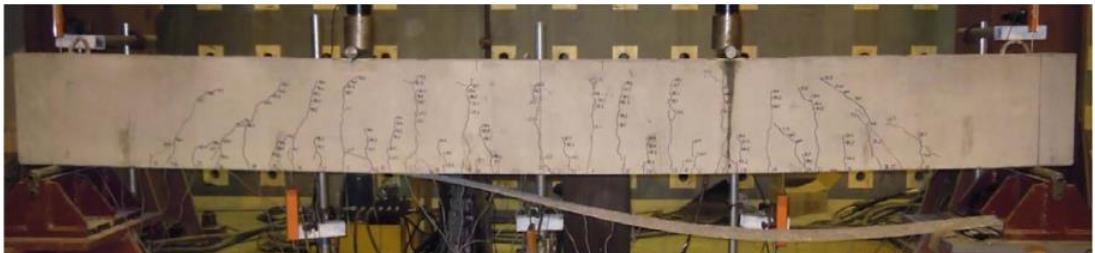
Fig.18 Failure mode of experiment [31]

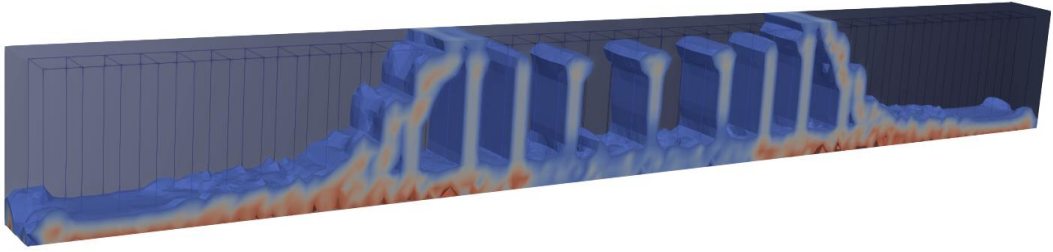

Fig.19 Failure mode obtained from simulation

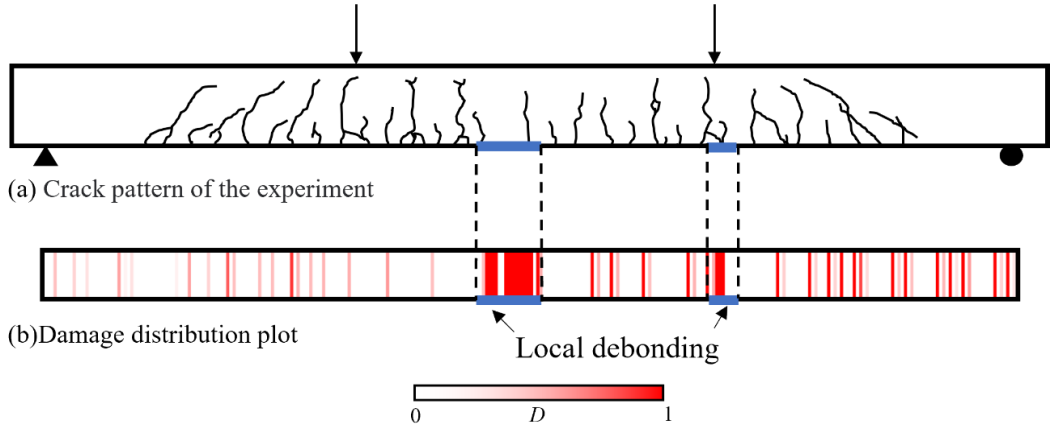

Fig.20 Crack pattern of the experiment and Damage distribution plot of cohesive element

Fig.20 displays the diagram of local debonding marks from the simulation. The result is obtained from the calculation step which represents the debonding initiation. The damage distribution map is derived using the average value along the thickness direction (as shown in Fig.20(b)). It is revealed that the numerically predicted zones of concentrated damage are close to the local debonding marked areas of experiment.

# 5. Conclusion

The interface parameters of ST-CZM exhibit significant differences compared to those of the traditional model. Due to the complexity of the problem, a primary challenge in employing the advanced CZM is calibrating the interface parameters. In this study, we propose a novel inverse identification framework utilizing the MIGA algorithm for optimizing the interface parameters. This framework considers both the structural response (i.e. the load-displacement curve) and the failure mode of the specimen (quantified by the debonding length) in objective function. The novel objective function is formulated by the least-squares norm. Through the DCB test, it is confirmed that the present framework can achieve precise interface parameter identification. Subsequently, the robustness and accuracy of the algorithm are verified. To further ensure that the optimized parameters can reproduce the failure mode of structures, the interface parameters of the four-point bending test are identified. The results demonstrate that the proposed framework for inverse identification remains insensitive to initial guesses and can accurately reproduce both the structural response and the failure mode of the experiment.

**CRediT authorship contribution statement**

**Tianxiang Shi:** Methodology, Investigation, Validation, Writing original draft. **Yongqiang Zhang:** Writing review and editing, Supervision, Project administration, Data curation.

**Declaration of competing interest**

The authors declare that they have no known competing financial interests or personal relationships that could have appeared to influence the work reported in this paper.

**Data availability**

Data will be made available on request.

**Acknowledgements**